# High-Performance Physics Simulations Using Multi-Core CPUs and GPGPUs in a Volunteer Computing Context

Kamran Karimi    Neil G. Dickson    Firas Hamze

D-Wave Systems Inc.
100-4401 Still Creek Drive
Burnaby, British Columbia
Canada, V5C 6G9
{kkarimi, ndickson, fhamze}@dwavesys.com

**Abstract**
This paper presents two conceptually simple methods for parallelizing a Parallel Tempering Monte Carlo simulation in a distributed volunteer computing context, where computers belonging to the general public are used. The first method uses conventional multi-threading. The second method uses CUDA, a graphics card computing system. Parallel Tempering is described, and challenges such as parallel random number generation and mapping of Monte Carlo chains to different threads are explained. While conventional multi-threading on CPUs is well-established, GPGPU programming techniques and technologies are still developing and present several challenges, such as the effective use of a relatively large number of threads. Having multiple chains in Parallel Tempering allows parallelization in a manner that is similar to the serial algorithm. Volunteer computing introduces important constraints to high performance computing, and we show that both versions of the application are able to adapt themselves to the varying and unpredictable computing resources of volunteers' computers, while leaving the machines responsive enough to use. We present experiments to show the scalable performance of these two approaches, and indicate that the efficiency of the methods increases with bigger problem sizes.

## 1. Introduction
Many fields of science and technology require vast computational resources. Simulation of physical systems is one notable example. Insufficient computing power can result in unfeasibly long running times or poor accuracy of results. Parallelizing such applications enables use of more processing power. However, parallelization of different applications may need different approaches in order to effectively use the processing units. In this paper, we parallelize a common Monte Carlo simulation technique known as Parallel Tempering Monte Carlo (PTMC) [8].

We focus on two technologies that enable parallelism: 1) multi-core Central Processing Units (CPU), and 2) streaming processors on Graphics Processing Units (GPU) [15]. In particular, we use NVIDIA's CUDA for GPU processing [15], though similar design principles may be applied to other GPU brands. Our primary design goals have been suitability for a volunteer computing environment, simplicity of the parallelization method, and low synchronization overhead. All of these goals have been achieved by leveraging the characteristics of our original (serial) algorithm.

The nature of Parallel Tempering Monte Carlo allows us to partition a large simulation into smaller, independent simulations, groups of which can be solved by threads on a CPU. The number of threads is chosen to be the same as the number of cores present in the system to avoid cache contention among the threads. GPGPUs allow for the execution of threads on a larger number of processing elements. Although these processing elements are typically much slower than those of a CPU, having a large number of threads may make it possible to surpass the performance of current multi-core CPUs.

In GPU programming, the creation of small, short-lived threads has been emphasized as a means towards achieving good speedup. An example would be multiplying two matrices, which can be done by dividing the task into threads, each of which computes one element of the target matrix. This low-level parallelism can be used in many algorithms and has been a significant source of GPU parallelization success. We show that, similar to CPU multi-threading, one can design a suitable GPU parallel algorithm that uses a coarser level of parallelization, running longer sequences of code on each GPU processor.

One matter that differs between multi-threaded CPU and GPU programming is memory structure. In a multi-threaded CPU application, all threads have access to a common memory space, so the primary challenge is synchronizing access to shared data structures. GPU programming, on the other hand, involves threads that run in a separate memory space from the main application, (which runs on the CPU). The implication is that the code and the data on which the GPU threads operate must be transferred to the GPU memory before any processing can be started, and the results of the computation must be copied back to the main memory of a host computer. This data transfer can result in poor performance and must be kept to a minimum. It is thus important to make sure that the time spent processing data on the GPU is long compared to data transfer times.

Another characteristic of parallel programming with GPUs is the ability to start a large number of threads with little overhead [15]. This is unlike traditional CPU threads, where each individual thread is treated as an entity independent of others, requiring separate resources such as stack memory, and whose creation and management (scheduling, keeping statistics of resource usage, etc.) are not cheap [15]. To absorb these costs, one option is to create as few CPU threads as possible and to make them run as long as possible. GPU threads, on the other hand, are cheaper to create and manage, since batches of GPU threads (blocks) are treated the same, so it is possible to create a large number of them and run them for shorter durations.

The specific application examined in this paper, called Adiabatic QUantum Algorithms (AQUA) [16], uses a Quantum Monte Carlo [3] algorithm to compute the minimum energy gap between two quantum states undergoing an adiabatic evolution. These simulations require substantial computational resources, and we need to solve many problem instances. We have employed the Berkeley Open Infrastructure for Network

Computing (BOINC) [2], to create the AQUA@home project and run the AQUA application on volunteers' computers in a massively distributed manner.

Resorting to volunteer computing entails serious considerations. An application that runs on a volunteer's computer cannot drain the computing resources, as that would make it hard for the volunteer to use his or her machine. It also cannot dynamically communicate with other computers. The problem of varying system resources and processing power on different computers must also be considered. Our algorithm has been designed with these constraints in mind, and the first constraint has had a large effect on the design of the GPU version. The problem is that using a GPU at full throttle makes the computer's user interface very slow or even non-responding, so caution must be used.

The rest of the paper is organized as follows. Section 2 briefly presents the PTMC algorithm and how its design naturally leads to parallel execution. Section 3 describes the method we used to parallelize random number generation similarly in both the CPU and GPU environments. Section 4 shows how our application was parallelized to use multiple CPU cores, while Section 5 explains the corresponding design for the GPU. We dedicate more space to the description of GPU parallelization because it is a less developed topic. Section 6 presents the results of performance and scalability tests. We conclude the paper in Section 7.

## 2. Parallel Tempering Monte Carlo

GPUs have been used to speed up QMC simulations before [1]. In those implementations, the parallelized components are primarily low-level matrix operations. Here, we present PTMC as an algorithm that is particularly well-suited for parallel execution. This variant allows a natural mapping of the algorithm to parallel execution units such as a CPU's cores or a GPU's streaming processors, enabling a high-level break down of the algorithm.

PTMC is based on the idea of running multiple Markov chains (sometimes referred to as replica) at the same time, each of which has as its stationary distribution that of the system at a different temperature. After a certain number of sweeps through all of the chains, i.e. flipping bits with the Metropolis algorithm according to the relevant temperatures [11], adjacent chains are probabilistically swapped. The probability of a chain swap is determined by the relative probability of the chains' states occurring at each other's temperature. In a non-parallel implementation of PTMC, the program would sequentially iterate through the set of chains to perform a sweep. After all the chains have been swept the required number of times, the program attempts to swap the chains. This cycle continues until the application's criteria for termination, such as a specified number of sweeps or satisfaction of a convergence test are met. More information about our implementation of Parallel Tempering comes in [7].

Each Monte Carlo chain can be swept independently of the others, so sweeping can be done in parallel. In our implementation, swapping chains can also be parallelized, because we consider only every two neighboring chains as possible swapping candidates, and then move to the next two chains. With *N* chains, up to *N*/2 swap operations could be done concurrently.

Selecting logical temperature values to achieve a desired minimum swap rate, even in the presence of first-order phase transitions, is described in [7]. After the process of chain generation, the number of chains and their temperature values are held constant during the PTMC simulation. In this paper we consider the chains to have been already generated.

**3. Parallel Random Number Generation**
PTMC simulations, like other stochastic methods, depend on a vast supply of (pseudo)random numbers, because both sweeping a chain and swapping chains require random numbers. It is very important to make sure that in a parallel environment, random numbers are generated in a fast and valid manner.

The original (serial) version of our application uses the well-known Mersenne-Twister algorithm [9]. The code for this algorithm is compact and contains simple integer operations. It relies on a global state vector and an index into that vector to generate random integers. This global set of variables makes it difficult to run the algorithm in parallel. Enforcing mutually exclusive access to this data structure through locks would cause poor performance, as a significant portion of time would be spent generating random numbers in serial.

A simple and effective solution to this problem is to create multiple state vectors, one for each CPU thread. Even though the code remains almost identical, each thread uses its own state to generate the next random number. Each chain's state has a different initial seed to prevent the problem of generating the same random numbers for different chains at every step. This simple solution allows us to maintain the same per-thread random number generation speed as in the single-thread case, with no compromise on the quality of the results.

On a CPU, each chain is swept by a single core, but as explained later, on a GPU, each chain is swept by a number of streaming processors. In both cases, a state vector is created for every execution thread that could be running in parallel with other execution threads to avoid the need for synchronization. As a result, on the CPU, a chain has a single random state, while on the GPU, each divisible part of a chain has its own random state.

Figures 1, 2, and 3 show how the original (serial), multi-threaded, and GPU versions initialize their Mersenne-Twister states, respectively. The Mersenne-Twister implementation itself is effectively identical between the three.

```
#define MT_STATE_LENGTH 624
static unsigned long randomState[MT_STATE_LENGTH];
static int randomIndex;
static void initializeRandom(unsigned long seed);
...
initializeRandom(seed);
```
Figure 1. Original (serial) Mersenne-Twister initialization.

```
#define MT_STATE_LENGTH 624
static void initializeRandom(unsigned long seed,
                             unsigned long* randomState,
                             int& randomIndex);
...
for (int chain=0;chain<numChains;chain++) {
    chains[chain].randomState = new unsigned long[MT_STATE_LENGTH];
    initializeRandom(startSeed+chain,
                     chains[chain].randomState,
                     chains[chain].randomIndex);
}
```
Figure 2. Multi-threaded Mersenne-Twister initialization. Each chain has its own state.

```
#define MT_STATE_LENGTH 624
__device__ unsigned long
     randomStates[MAX_RAND_CHAINS][MAX_RAND_THREADS][MT_STATE_LENGTH];
__device__ int randomIndices[MAX_RAND_CHAINS][MAX_RAND_THREADS];
__global__ void cuda_init_genrand(unsigned long seed,
                                  unsigned long* randomState,
                                  int* randomIndex);
...
int thread = threadIdx.x;
int chain = blockIdx.x;
initializeRandom(startSeed + chain*threadDim.x + thread,
                 randomStates[chain][thread],
                 &randomIndices[chain][thread]);
```
Figure 3. GPU Mersenne-Twister initialization. Each chain can have multiple threads.

### 4. Multi-Threading on Multi-Core CPUs

We parallelized the sweep and swap parts of the Parallel Tempering code using OpenMP [5]. By default, OpenMP creates the same number of threads as the number of available cores in the computer. Since this number is currently less than the number of chains (typically 4 or 8 cores vs. 25 to 200 chains), each thread is responsible for sweeping multiple chains.

An alternative to one thread per core is to have one thread per chain. For a small number of chains, this scheme can have advantages in terms of less time spent in waiting on barriers. However, miscellaneous timing tests suggested that even for a relatively small number of chains, the performance was worse. We suspect this to be because of increased CPU cache misses and thread management overhead.

When parallelizing a loop with *I* iterations, where each iteration of the loop can be done independently, OpenMP assigns *I*/*N* iterations (or one of the nearest integers thereof) to each of the *N* threads. Its default behavior is to start the worker threads when the loop is encountered, and wait for all threads to finish their iterations before continuing. Figure 4 shows the outline of our first attempt at parallelizing the Parallel Tempering algorithm. Unrelated code is omitted for brevity.

```
while (!shouldTerminate()) {
    #pragma omp parallel for
    for (int chain=0; chain<numChains; chain++) {
        for (int sweep=0; sweep<numSweepsPerSwap; sweep++) {
            runMetropolis(chains[chain]); } }
    doSwap(chains);
    // Do any meaurements, gather statistics, etc. here
}
```

Figure 4. Outline of Parallel Tempering with OpenMP

In Figure 4, *chains* contains a pointer to the Parallel Tempering chain data. The outer for-loop is run in parallel. Each thread is thus responsible for running the inner loop for one or more *chain* values. To indicate that OpenMP should parallelize the for-loop, one adds the #pragma statement. No data that are modified within the for-loop are shared between the threads, so they can run with no need for synchronization. In this code the threads join before the doSwap() method.

In our simulation, different chains are at different effective temperatures. Chains at higher temperatures have more bit flips and therefore require more work to update the state after a flip. OpenMP's default assignment of chains to threads does not result in a balanced load distribution, because it assigns successive chains to each thread, resulting in some threads with many easy (cold) chains and some threads with many hard (hot) chains. Therefore, we form a work pool of chains, and start by assigning harder chains to each available thread in a first-come, first-served manner. Easier chains are assigned last. We observe the number of active cores to determine the efficiency of our work-assignment method, and the described method results in all the cores being occupied most of the time.

Further improvement in CPU utilization was achieved by keeping the threads active throughout the simulation. However, barriers must then be specified explicitly to synchronize the threads between phases. Swapping was also parallelized, so it required the current thread index and the total number of threads.

Figure 5 shows the relevant code that implements the above improvements.

```
static volatile int chain; // The workpool index
#pragma omp parallel {
    int thread = omp_get_thread_num();
    do {
        // Sweeping through the chains selected from the workpool
        if (thread==0) chain = numThreads;
        #pragma omp barrier
        int myChain = thread;
        while (myChain < numChains) {
            for (int sweep=0; sweep<numSweepsPerSwap; sweep++) {
                runMetropolis(chains[myChain]); }
            // Get the next chain without colliding with other threads
            #pragma omp critical
            {myChain = chain++;}
        }
        #pragma omp barrier
        doSwap(chains,thread,numThreads); // Swapping
        // Do any meaurements, gather statistics, etc. here
    } while (!shouldTerminate());
}
```

Figure 5. Keeping threads active; sweeping loop with workpool.

AQUA@home is the first BOINC project to support CPU multi-threading. With BOINC, the user can determine how many cores are to be used, so BOINC provides the application with the number of threads that should be created. BOINC usually runs an application at a below-normal priority to make sure the computer remains responsive. However, this priority level is not respected for newly created OpenMP threads. AQUA explicitly sets the priority of all the newly created OpenMP threads to the same value as determined by BOINC. As a result, even though AQUA has a high CPU utilization rate (about 94% on 8-core computers), we have observed that the above method of parallelizing the PTMC algorithm allows volunteers to continue to use and interact with their machines.

**5. Parallelization on GPGPUs**

GPU programming involves more overhead than CPU programming because of the need to transfer data to and from the graphics card. Once the data are in place on the card, many threads can be started with little overhead. The code that is invoked on the GPU and runs in many threads is called the kernel. At the end of the computation, the results should be copied back to the main memory of the computer.

A GPU contains a number of Multi-Processors (MP). Each of these in turn has a certain number of processing elements, so parallelism on a GPU is both at the MP and at the processing element level. These two levels of parallelism are reflected in adapting the PTMC method to run on a GPU. The first level involves mapping the chains to the MPs. This is similar to the multi-threading case, where each chain is swept by a thread. Each MP thus plays a role similar to a CPU core. The second level of parallelism is using the streaming processors in each MP to parallelize the sweep of individual chains.

Significant factors in obtaining good speedup with a GPU include reducing the data transfer overhead, reducing the required GPU resources per thread so that more threads can be started, and keeping all of the MPs busy.

A Monte Carlo sweep usually involves changing the state of the chain, and the implied write operations need synchronization if performed in parallel. However, chains in our simulations have regions that can be updated independently, since they are laid out such that a change in one region does not directly affect another. In our case, one group of many independent regions (64 in results below) covers half of the chain, and a second group covers the other half. The first group can be swept in parallel without synchronization. After that the threads are synchronized before continuing with the second group of regions.

Each thread block sweeps different parts of the same chain, and some data common to the chain are placed in the shared memory in order to be accessed faster than if they were in the device's global memory. An added benefit to this scheme is that more registers are freed to be used for other purposes. However, in our case managing the shared memory block was difficult, because each chain has more data than could fit into the shared memory.

To keep all the streaming processors busy, we assign multiple chains to each MP, a method we call chain packing. This is similar to the CPU case where multiple chains were assigned to the same core by OpenMP. The main limiting factors in the GPU case are the number of registers available per MP and the maximum number of threads per MP. Each MP in current NVIDIA GPUs can take up to 512 threads, giving us a maximum of 8 chains per MP (assuming 64 threads per chain). All threads in an MP share a pool of registers, so the maximum number of used registers must not exceed this maximum. On older cards with 8,192 registers, our code can run 4 chains per MP, while with cards with 16,384, we can run the maximum possible 8 chains per block (these numbers may change for future GPUs). For example, with 30 MPs we can run 120 or 240 chains in parallel, or more if fewer threads per chain are used. This high number explains the attractiveness of GPU parallelization, because in spite of streaming processors being slower than a typical CPU, their sheer number can result in an overall speedup.

On the other hand, with a small number of chains, packing the MPs to the maximum may result in some idle MPs. For this reason, we stop packing chains in the MPs as soon as we see that some MPs are not being utilized.

Under BOINC, it became apparent that volunteers expect as little CPU activity as possible from a GPU application. One reason is that many volunteers use their computers to run CPU-heavy BOINC applications at the same time as running GPU applications. However, loading the GPU with long-running kernels makes the system sluggish and eventually freezes the computer's user interface, making it unusable for the volunteer. The responsiveness of the computer can be adjusted by changing the ratio of the time spent executing code on the GPU vs. on the CPU. We chose for swapping to be done on the CPU, which allows us to change the GPU vs. CPU running time ratio by changing the number of sweeps between each swap operation. The application thus alternates between using the GPU to sweep the chains and then using the CPU to swap them.

Data need to be transferred to the GPU before sweeping, and must be moved back to the host computer before swapping. As we mentioned before, copying data to and from the GPU is a concern. We increase the number of sweeps in between swap phases to absorb the cost of the data copying, but not so much as to make the computer unresponsive.

The issue of the required floating point precision of the application must also be addressed. Recent graphics cards support double-precision floating point operations, but they may run slower than single-precision operations [15]. In our application, the sweeping phase performs a mix of integer and floating-point operations. Real-valued data that are processed on the GPU are represented as single-precision floating-point numbers. The corresponding code in the CPU-only version also uses single-precision. In both versions, the results are accumulated on the CPU in double-precision numbers in order to minimize building up round-off error.

Figure 6 shows the code for running the same loop as in Figure 4 in parallel with a GPU. `cuda_runMetropolis`(numSweepsPerSwap) performs the specified number of sweeps on the GPU before returning control to the CPU.

```
cuda_alloc_mem(chains); // Allocate memory on the graphics card
while (!shouldTerminate()) {
  cuda_copy_chains_to_device(chains); // Copy data to the graphics card
  cuda_runMetropolis(numSweepsPerSwap);
  cuda_copy_chains_to_host(chains);   // Copy data back to main memory
  doSwap(chains);
  // Do any meaurements, gather statistics, etc. here
}
cuda_free_mem();         // Free memory on the graphics card
```

Figure 6. Main loop of a GPU Parallel Tempering algorithm.

Figure 7 shows how the code to run on the GPU is prepared. First, the number of chains assigned to each MP is calculated. The kernel is then invoked and a test is performed to see whether the kernel was successfully started given its configuration. If a failure due to a lack of resources is detected, the code reduces the number of packed chains, thus creating more blocks, and reducing the needed resources such as the number of registers. The kernel is then tried again. In this way, the code can adapt itself to the capabilities of the GPU on which it is running.

```
//cuda_MP_count contains the number of MPs on the device
static int block_size; // Number of threads in each block
static int packed_chains; // number of chains in each MP
static int num_blocks; // number of blocks in each MP
static int num_nodes;  // numbr of nodes assigned to each thread
static int last_num_chains = 0; // the previous number of chains

void cuda_singleSiteMetropolis(/* arguments */) {
 cudaError err;

 last_num_chains = num_PT_chains; block_size = 32; packed_chains = 1;
 num_blocks = num_PT_chains / packed_chains +
           ((num_PT_chains % packed_chains) == 0? 0 : 1);

 // pack the chains but stop when MPs start going unused.
 while (packed_chains * block_size < MAX_CUDA_BLOCK_THREADS &&
           num_blocks > cuda_MP_count) {
   packed_chains *= 2;
   num_blocks = num_PT_chains / packed_chains +
                   ((num_PT_chains % packed_chains) == 0? 0 : 1);
 }
 int kernel_success = 0;
 while(kernel_success == 0) {
   kernel_success = 1; // was the kernel started successfully?
   dim3 gridDim(num_blocks); dim3 blockDim(block_size, packed_chains);
   cuda_runMetropolis <<<gridDim, blockDim>>> (/* arguments */);
   //kernel call
   cudaThreadSynchronize(); //make sure all threads are finished
   err = cudaGetLastError(); // did anything go wrong?
   if(err != cudaSuccess) {
    if(err == 7 && packed_chains>1) {//not enough resources (registers)
         packed_chains /= 2;
         num_blocks = num_PT_chains / packed_chains +
                     ((num_PT_chains % packed_chains) == 0? 0 : 1);
         kernel_success = 0; }
    else { /* Error starting the kernel. Exit the program */ }
   }
  }
 }
}
```

Figure 7. Choosing the number of threads per MP and the number of blocks.

In the above code, each MP will have *num_threads_per_block* × *packed_chains* threads. There will be *num_blocks* number of such groups of threads. If this number is less than the number of MPs, they will all run in parallel. However, if *num_blocks* exceeds the number of MPs, then the MPs will take turns sweeping the chains.

**6. Performance and Scaling Experiments**
In this section, we investigate the performance and scaling of our CPU-only and CUDA applications. It is difficult to make a fair comparison between an application that runs only on a CPU and the same app that uses a GPU, because the comparison results will change if the employed CPU or GPU are changed. For this reason we will not perform any direct comparisons between the two implementations.

We used 6 different problems for our experiments, ranging from 8 to 96 qubits, as listed in Table 1. In order to approximate the quantum equilibrium properties of a quantum system, we connect together many copies (128 in this case) of the corresponding classical system [14]. This larger classical system is then simulated in each chain. The total number of variables in the simulation is the product of the number of qubits, the number of copies in each chain, and the number of chains. For example, to simulate an 128 copies of an 8-qubit quantum system with 27 chains we have $128 \times 8 \times 27 = 27{,}648$ classical variables to manage. The rapid increase in the number of variables required to simulate a quantum system necessitates good simulation performance.

| Qubits | Chains | Total number of variables |
|---|---|---|
| **8** | **27** | 27,648 |
| **16** | **34** | 69,632 |
| **32** | **37** | 151,552 |
| **48** | **57** | 350,208 |
| **72** | **71** | 654,336 |
| **96** | **111** | 1,363,968 |

Table 1. The size of problems used in the experiments

To measure the effectiveness of our multi-threaded CPU code, we ran the AQUA application on 32-qubit and 96-qubit Ising model [6] problems with 1, 2, 4, 6, and 8 threads, on a Mac Pro computer with two 2.8 GHz Intel Quad Core Xeon processors (total of 8 cores) and 2 GB of 800 MHz DDR2 memory, running Mac OS 10.5. We performed 200,000 sweeps per run. Each run was repeated 10 times to obtain reliable average running times. The amount of time needed to perform the parallel tempering, which involves no I/O, is noted in Table 2. For completeness, we also note the total running time of the application, which involves some I/O operations. The GPU was not used in this case.

| Qubits | Threads | PT Time | Std. Dev. | Total Time | Std. Dev. |
|---|---|---|---|---|---|
| **32** | 1 | 2,517.01 | 0.25 | 2,625.08 | 0.26 |
|  | 2 | 1,300.82 | 14.36 | 1,355.10 | 14.44 |
|  | 4 | 691.93 | 0.15 | 719.34 | 0.16 |
|  | 6 | 480.83 | 0.16 | 499.20 | 0.16 |
|  | 8 | 380.43 | 0.18 | 394.32 | 0.18 |
| **96** | 1 | 9,190.18 | 1.62 | 9,598.28 | 1.63 |
|  | 2 | 4,695.24 | 10.02 | 4,901.03 | 10.06 |
|  | 4 | 2,398.06 | 1.46 | 2,501.24 | 1.47 |
|  | 6 | 1,653.01 | 0.18 | 1,722.86 | 0.18 |
|  | 8 | 1,273.01 | 0.12 | 1,324.95 | 0.13 |

Table 2. Multi-threading performance for 32- and 96-qubit problems.

Figure 8 shows the speedup obtained using different number of cores. Also shown is the theoretical limit of linear speedup. We see a good scaling of the running time with the number of threads, which improves with larger problem sizes, since larger problems have more chains, with each chain needing the sweeping of more variables. There are thus more independent tasks that can be run in parallel between barriers.

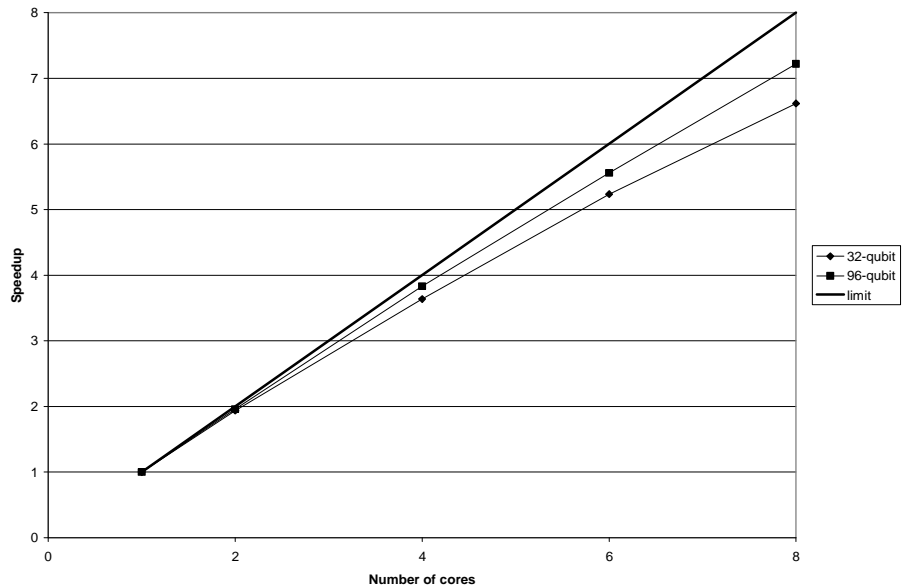

Figure 8. Speedup vs. number of cores for the multi-threaded application.

Our GPU experiments were performed on a mid-range NVIDIA GeForce GTX 260 graphics card in a dual-core PC running Windows XP. In this case we packed the card to the maximum possible, and solved problems that differed in size, to observe performance of the GPU as a function of the problem size. The 32- and 96-qubit problems are the same as in the multi-threaded case. We again performed 200,000 sweeps per run. The results come in Table 3. Only one CPU thread was employed.

| Qubits | PT Time | Std. Dev. | Total Time | Std. Dev. |
|---|---|---|---|---|
| 8 | 274.79 | 0.41 | 276.39 | 0.42 |
| 16 | 575.67 | 0.56 | 579.63 | 0.56 |
| 32 | 1,138.33 | 0.49 | 1,146.72 | 0.49 |
| 48 | 2,621.03 | 3.51 | 2,641.30 | 3.51 |
| 72 | 4,849.05 | 2.84 | 4,892.47 | 2.84 |
| 96 | 10,107.71 | 2.47 | 10,213.52 | 2.47 |

Table 3. Running time for problems of different size for the GPU application in seconds

Here again, we see an improvement in performance as the problem size increases. One reason is that with bigger problems, proportionally more time is spent sweeping the chains than copying data over to the host computer for swapping. Figure 9 shows this relative performance increase in terms of the number of variables per unit time spent in the parallel tempering part of the code. In this log-linear plot, the number of variables in the problem is displayed next to the corresponding data point. The data points are obtained by dividing the total number of variables by the parallel tempering time. The observed gain in performance saturates by 654,336 variables (72 qubits), which is where the maximum number of the GPU streaming processors are utilized.

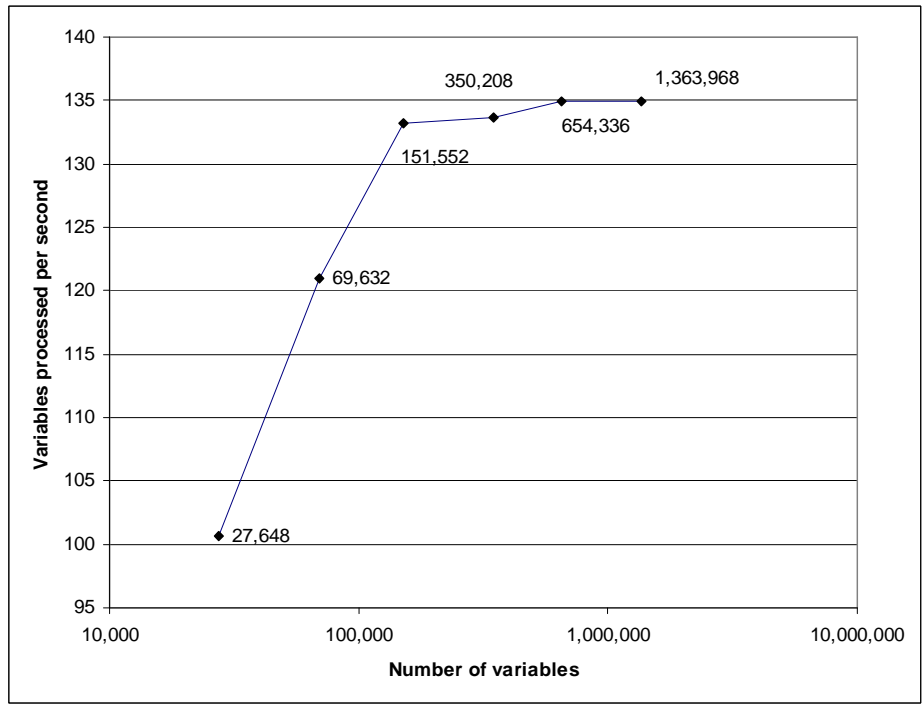

Figure 9. GPU performance in relation to problem size

The main reason for the relatively low performance of the GPU version compared to the CPU version is that high GPU utilization has a dramatic effect on the responsiveness of the computer. We consciously refrained from putting maximum load on the GPU, to allow the volunteers to be able to use their computers while our application is running. There are also algorithmic issues that rob performance: sweeping each chain involves randomly deciding whether or not to flip each bit. On a GPU MP, even if only one streaming processor decides to flip a bit, all threads on that MP must wait an amount of time equal to executing that flip, since they share an instruction pointer. The resulting performance is as if most chains always flip all bits.

The main cost of data transfer to the GPU is the latency time required to start a transfer, while the data transfer itself is relatively fast. We observed improvements in execution speed by coalescing data structures into array forms, so the data can be transferred in fewer operations.

**7. Concluding Remarks**
We introduced two methods to parallelize a Parallel Tempering Monte Carlo algorithm on multi-core CPUs and on GPGPUs. The methods are conceptually simple, because they assign independent data processing to CPU threads (PTMC chains) or GPU threads (parts of PTMC chains). The resulting software is meant to run on volunteer computers belonging to members of the general public. It adapts itself to the available hardware, and employs the computing resources in a manner that leaves the computer on which it runs usable.

We presented experimental data showing that the parallel algorithms are scalable within the ranges of the currently available computational resources. The constraint of leaving the computer with an appropriate level of usability makes our code slower than if it could use all of the available computing resources, particularly for the GPU application. However, it has the advantage of being able to run on thousands of volunteer computers, making up for lowered efficiency. The resulting application, AQUA, successfully solves problems that require vast computing resources.

**Acknowledgement**
We would like to thank Geordie Rose for his help with this project. We are grateful to the developers of BOINC, at the University of California at Berkeley, for their help with the AQUA@home project. We thank all of the volunteers of AQUA@home, without whom this project would not be successful.

**References**
[1] Anderson, A.G., Goddard, W.A., Schröder, P., Quantum Monte Carlo on Graphical Processing Units, *Computer Physics Communications*, Volume 177, Issue 3, 2007.
[2] Anderson, D.P., BOINC: A System for Public-Resource Computing and Storage, *The fifth IEEE/ACM International Workshop on Grid Computing*, pp. 365-372, Pittsburgh, PA, USA, 2004.
[3] Anderson, J.B., *Quantum Monte Carlo: Origins, Development, Applications*, Oxford University Press US, 2007
[4] Berg, B.A., Markov *Chain Monte Carlo Simulations and Their Statistical Analysis*, World Scientific, 2004
[5] Chapman, B., Jost, G. and van der Pas, R., *Using OpenMP: Portable Shared Memory Parallel Programming*, The MIT Press, 2007
[6] Fischer, K. H. and Hertz, J. A. *Spin Glasses*, Canbridge University Press, 1993.
[7] Hamze, F., Dickson, N., Karimi, K., Robust Parameter Selection for Parallel Tempering, *International Journal of Modern Physics C*, accepted.
[8] Hukushima, K. and Nemoto, K.,' Exchange Monte Carlo Method and Application to Spin Glass Simulations*, Journal of the Physical Society of Japan 65*, 1996, pp. 1604-1608.
[9] Matsumoto, M. and Nishimura, T., Mersenne Twister: A 623-Dimensionally Equidistributed Uniform Pseudo-Random Number Generator, *ACM Transactions on Modeling and Computer Simulation,* Vol. 8, No. 1, January 1998, pp 3--30.
[10] Neven, H., Denchev, V.S., Rose, G., Macready, W.G., Training a Binary Classifier with the Quantum Adiabatic Algorithm, *Neural Information Processing Systems (NIPS) workshop on Optimization for Machine Learning*, 2008.

[11] Robert C.P., and Casella, G., *Monte Carlo Statistical Methods* , Springer, 2005.
[12] Rubinstein, R. Y.; Kroese, D. P., *Simulation and the Monte Carlo Method* (2nd ed.), John Wiley & Sons, 2007.
[13]Wu, J., *Distributed System Design*, CRC Press, 1999.
[14] Young, A.P., Knysh, S. , Smelyanskiy, V. N., Size dependence of the minimum excitation gap in the Quantum Adiabatic Algorithm, *Physical Review Letters 101, 170503*, 2008
[15] Kirk, D. and Hwu, W., *Programming Massively Parallel Processors: A Hands-on Approach*, Morgan Kaufmann, 2010.
[16] AQUA's home page at http://aqua.dwavesys.com